\documentstyle[prl,aps,tighten,multicol,epsf]{revtex}

\newcommand{\one}  {\mbox{$\openone$}} 
\newcommand{\be}   {\begin{equation}} 
\newcommand{\ee}   {\end{equation}} 
\newcommand{\ba}   {\begin{eqnarray}} 
\newcommand{\ea}   {\end{eqnarray}}

\begin{document} 
 
\title{Conductance fluctuations and weak localization in 
       chaotic quantum dots}

\author{E. R. P. Alves and C. H. Lewenkopf} 
 
\address{Instituto de F\'{\i}sica,  
	   Universidade do Estado do Rio de Janeiro,\\  
	   R. S\~ao Francisco Xavier, 524,  
	   20559-900 Rio de Janeiro, Brazil} 
 
\date{\today} 
 
\maketitle 
  
%%%%%%%%%%%%%%%%%%%%%%%%%%%%%%%%%%%%%%%%%%%%%%%%%%%%%%%%%%%%%%%%%%%%%%%%%%% 
\begin{abstract} 
We study the conductance statistical features of ballistic electrons 
flowing through a chaotic quantum dot. We show how the temperature 
affects the universal conductance fluctuations by analyzing the influence 
of dephasing and thermal smearing. This leads us to two main findings.
First, we show that the energy correlations in the transmission, which
were overlooked so far, are important for calculating the variance and 
higher moments of the conductance.
Second, we show that there is an ambiguity in the method of determination 
of the dephasing rate from the size of the of the weak localization.
We find that the dephasing times obtained at low temperatures from quantum
dots are underestimated.
\end{abstract} 
%%%%%%%%%%%%%%%%%%%%%%%%%%%%%%%%%%%%%%%%%%%%%%%%%%%%%%%%%%%%%%%%%%%%%%%%%%% 
 
 
% 03.65.Sq  Semiclassical theories and applications 
% 03.65.Yz  Decoherence; open systems; quantum statistical methods 
% 05.45.Ac  Low-dimensional chaos 
% 05.45.Mt  Semiclassical chaos ("quantum chaos") 
% 73.21.La  Quantum dots 
% 73.23.-b  Mesoscopic systems 
% 73.23.Ad  Ballistic transport 
%--------------------------------------------------------------------------  
 
\begin{multicols}{2} 
% now the plain text 

A very striking experimental evidence of universal statistical behavior 
due to quantum coherence and complexity in electronic ballistic transport 
was recently reported by Huibers and collaborators \cite{Huibers98a}. 
They measured the conductance $G$ through a chaotic quantum dot at small
bias and low temperatures as a function of an applied magnetic field  
and the quantum dot shape.  
For such devices, where the quantum coherence length $\ell_\phi$ and the 
system size $L$ are such that $\ell_\phi \gg L$, the conductance is 
expected to parametrically display mesoscopic fluctuations. 
\cite{Altshuler95} 
To characterize the latter, due to the system complexity, a detailed 
microscopic theory is neither feasible nor practical.
Hence, the indicated theoretical approach should be statistical and 
tailormade to give the experimental accessible statistical measures 
such as the conductance distribution $P(G)$, conductance autocorrelation 
functions, etc.. 
For ballistic chaotic quantum dots such approach is provided by the  
random matrix theory (RMT). 
Indeed, the agreement between the conductance distributions $P(G)$ 
obtained in Ref. \onlinecite{Huibers98a} and the corresponding 
stochastic theory, turned this experiment into a paradigm of the 
statistical approach. \cite{Beenakker97,Alhassid00} 
 
Early experiments \cite{Chang94,Chan95} revealed an unexpected aspect 
to that systems, namely that even at low temperatures the conductance 
fluctuations significantly deviate from the predictions of the simplest 
random matrix models \cite{Baranger94,Jalabert94}. 
More specifically, we are referring to the suppression of the weak 
localization peak, which represents the first quantum correction to the 
classical picture, and to the conductance variance $\mbox{var}(G)$. 
The early works were improved and converged to the understanding that 
even a small loss in quantum coherence \cite{Buttiker86} affects 
dramatically the statistical observables \cite{Huibers98a}.
At the quantitative level, some features of the experimental data
still remain unexplained.

The main findings presented in this letter are two-fold. First, using an 
alternative statistical approach we explain the discrepancy between
theory and experiment in Ref. \onlinecite{Huibers98a} for var($G$).
This result has important consequences for recent predictions of var$(G)$ 
in similar systems. \cite{outros}
Second, we show that there is an ambiguity in the way the dephasing  
rates are extracted from the weak localization experimental data in
open chaotic quantum dots so far. \cite{Huibers98b,Huibers99,Pivin99}
Within our statistical model we propose a different method, which 
indicates that the dephasing rates quoted in the literature 
\cite{Huibers99} are overestimated. 
 
The conductance $G = (e^2/h)g$ through a two-lead quantum dot is related to  
the transmission, and hence to the $S$ matrix, by the Landauer formula 
\be 
\label{eq:Landauer} 
g(E, X) \equiv T_{21}(E,X)  
        = \sum_{{a \in 1}\atop{b \in 2}} |S_{ba} (E,X)|^2 .
\ee 
Here $g$ is the dimensionless conductance, $T_{21}$ is the transmission 
of an electron scattered from the incoming lead 1 to the outgoing lead 2,  
and the labels of the corresponding scattering matrix $S$ indicate the  
open channels located at each lead. 
$X$ is a generic parameter such as a gate voltage, which shapes the dot, 
or an external applied magnetic field $B$. 
The applicability of the Landauer formula assumes full quantum coherent  
transport. 
 
Thermal effects modify  Eq.\ (\ref{eq:Landauer}) in different ways.
First, and most interesting, by increasing the temperature the  
dynamics in the dot changes, making the coherent single-particle 
description of the process less realistic. The rich physics involved 
attracted a lot of attention and a lively debate
lately \cite{experiments,Altshuler}. 
One way to include such dephasing processes in the theory 
is provided by the B\"uttikker phenomenological model \cite{Buttiker86}.
This approach is remarkably successful and its use became customary  
in the treatment of conductance fluctuation in chaotic dots.
\cite{Baranger95,Brouwer95,Brouwer97} 
It introduces a ficticious  voltage probe lead $\phi$,  through which 
there is no net current flow, but allows for electrons to randomize their 
phases at the reservoir $\phi$. 
As a result the dimensionless conductance reads 
\be 
\label{eq:Buettikker} 
g_\phi(E,X) = T_{21} + \frac{T_{2\phi}T_{\phi1}}{T_{1\phi}+T_{2\phi}} \;,
\ee 
where the arguments $E$ and $X$ are implicit to $T$.
 
Temperature also affects the conductance in another (rather trivial) manner: 
the electrons flowing through a quantum dot are thermally distributed, 
yielding
\begin{equation} 
\label{eq:gtemperature} 
G(\mu, X) = \frac{e^2}{h} \int\! dE\, g_\phi (E, X)  
            \left(-\frac{\partial f_\mu}{\partial E}\right) \,, 
\end{equation} 
where $f_\mu \equiv \{1+ \exp[(E-\mu)/T]\}^{-1}$ is the Fermi distribution 
function and $\mu$ is the chemical potential of the dot.
It should be emphasized that in our notation $g_\phi$ accounts 
solely for dephasing, while $G$ is affected both by dephasing and the 
smearing of the Fermi surface.
 
Let us consider the simplest statistical measures of $P(G)$, namely
the mean conductance $\langle G \rangle$ and its variance var$(G)$. 
In experiments $\langle G \rangle$ is obtained by varying $\mu$ and/or 
$X$, whereas in theory one takes a suitable ensemble averaging  
over $g_\phi$. 
Actually, from Eq.\ (\ref{eq:gtemperature}) it is evident that  
$\langle G \rangle = (e^2/h) \langle g_\phi \rangle$. 
The inspection of the conductance autocorrelation function 
\be 
C(x) = \langle G(\mu, X^+)G(\mu, X^-)\rangle - \langle G (\mu, X)\rangle^2 
\ee 
where $X^\pm = X \pm x/2$, reveals that the relation between var($G)[ = C(0)]$ 
and var$(g_\phi)$ is obtained from \cite{Efetov95} 
\be 
\label{eq:temperatureconvolution} 
C(x) = \left(\frac{e^2 T}{h}\right)^2 \!\!\!\!\int_{-\infty}^{\infty}\!\! d\omega 
      \,c_\phi(\omega, x)
      \frac{d}{dT} \left[ 2T \,\mbox{sinh} \!\left(\frac{\omega}{2T}\right)\right]^{-2} 
\ee 
where $c_\phi(\omega, x)$ is the dimensionless conductance  autocorrelation function defined by 
\be 
c_\phi(\omega, x) = \left\langle g_\phi(E^+, X^+) 
                            g_\phi(E^-, X^-)\right\rangle 
               -\langle g_\phi \rangle^2 
\ee 
with $E^\pm = E \pm \omega/2$.

So far all theoretical studies  
\cite{Baranger94,Jalabert94,Baranger95,Brouwer95,Brouwer97} aiming to  
describe $P(G)$ used the information theoretical approach.  
More specifically, one finds the ensemble of $S$ matrices that maximizes  
the information entropy and fulfills the symmetries and other constraints
of the physical system under consideration.
This procedure is very amenable for the analytical calculation of $P(g_\phi)$  
(at fixed $E$ and $X$) but has the limitation of lacking parametric  
correlations (neither for $E$ nor for $X$) between members of the ensemble. 
This ingredient is of central importance in obtaining the variance and higher 
moments of $G$ as indicated by Eq.\ (\ref{eq:temperatureconvolution}). 
To circumvent this problem Ref.\onlinecite{Huibers98a} introduced an
heuristic procedure of smearing $P(G)$, which underestimates 
$\mbox{var}(G)_{\beta=2}/\mbox{var}(G)_{\beta=1}$.
 
We use the Hamiltonian approach to the statistical $S$-matrix instead 
\cite{Verbaarschot85}. Both frameworks are equivalent for the
calculation of var($g_\phi$) \cite{Lewenkopf91}, but not for var$(G)$.
The resonant $S$-matrix is given by  
\be 
S(E, X) = \one - 2\pi i W^\dagger \frac{1}{E - H(X) + i \pi W W^\dagger} W 
\ee 
where $H$ is taken as a member of the Gaussian orthogonal (unitary) 
ensemble for systems where time-reversal symmetry is manifest (absent). 
For simplicity we take the case of $N$ open channels at each lead.
Since the $H$ matrix, of dimension $M$, is statistically invariant 
under orthogonal ($\beta=1$) or unitary ($\beta=2$) transformations, 
the statistical properties of $S$ depend only on the mean resonance 
spacing $\Delta$ determined by $H$ and on $W$, the coupling matrix elements 
between resonances and channels.
Those enter the theory through $y_c = \pi^2(W^\dagger W)_{cc}
/(M\Delta)$ contained in the so called sticking coefficients $P_c 
= 4y_c/(1 + y_c)^2$.
The later quantify the transmission through a given channels $c$, being 
maximal for $P_c=1$. By assuming the channels to be equivalent we can drop
the index $c$.
For open quantum dots the transmission is large and consequently it 
is assumed that $P\approx 1$.
In addition, 
we consider $N_\phi$ open channels at the voltage probe lead, each of them
with a sticking coefficient $p$. The loss of phase coherence is modeled by 
the single parameter $P_\phi = p N_\phi$, with $N_\phi\gg 1$ \cite{Brouwer97}.
The later can be converted in a dephasing width $\Gamma_\phi=\Delta P_\phi
/2\pi$, from which the dephasing time $\tau_\phi = \hbar/\Gamma_\phi$ is
extracted.

In this approach the parametric correlations are automatically taken into 
account, but due to technical reasons it is very difficult to proceed
analytically, unless $N\gg 1$ and $\beta=2$ \cite{Efetov95}. 
On the other hand, numerical simulations can be implemented 
straightforwardly. 
For each realization of $H$ we invert the propagator and compute 
$S$ for values close to the center of the band, $E=0$, where the 
$\Delta$ is approximately constant. The dimension of $H$ 
was fixed to be $M=200$, taken as a compromise between having 
a reasonable wide energy window to work with and not slowing too  
much the computation.  
For each case under consideration we obtained good statistics 
for $P(g_\phi)$ and $c_\phi(\omega, x)$ with $10^4$ realizations.  

\begin{figure} 
\setlength{\unitlength}{1mm} 
\begin{picture}(90,90)(0,0) 
\put(0,-22){\epsfxsize=8.cm\epsfbox{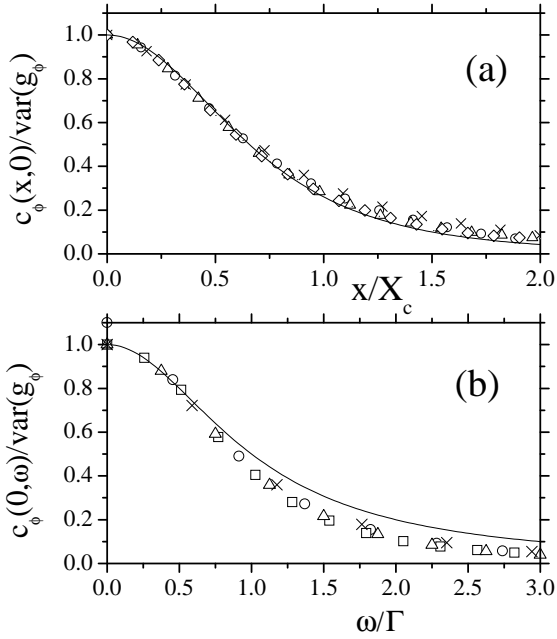}} 
\end{picture} 
\caption{Ratio between the dimensionless conductance autocorrelation 
function $c_\phi(x,\omega)$ and its variance $c_\phi(0,0)$ as a function 
of (a) $x/X_c$ with $\omega=0$ and (b) $\omega/\Gamma$ with $x=0$. 
The solid line stands for the result of Eq.\
(\ref{eq:naive}), where as the numerical simulations for $P_\phi = 1.0,
2.0, 3.0,$ and 4.0 are represented by the symbols $\times, \Box, \circ,$
and $\triangle$ respectively.}
\label{fig:scaling+dephasing} 
\end{figure}

We find that for the case of experimental interest, $N=1$ and 
$P_\phi\neq 0$, the numerically computed dimensionless 
autocorrelation function $c_\phi(\omega, x)$ is quite similar 
to the one obtained in the semiclassical regime ($N\gg 1$) and 
$P_\phi=0$ \cite{Efetov95,Baranger93,Vallejos01}, namely 
\be 
\label{eq:naive}
c_\phi(\omega, x) \approx \frac{\mbox{var}(g_\phi)}{[1+ (x/X_c)^2]^2 + 
(\omega/\Gamma)^2} .
\ee 
Our results, shown for the $\beta=1$ case in Fig. \ref{fig:scaling+dephasing}
($\beta =2$ gives essentially the same agreement), scale according to
\be
\Gamma = \Gamma_0 + \frac{\Gamma_\phi}{\beta} 
= \frac{\Delta}{2\pi}\left(2N P + \frac{P_\phi}{\beta}\right)
\ee
and $X_c = \kappa \sqrt{2NP + P_\phi}$, where $\kappa$ is system specific and
depends on the kind of parametric variation.
It is worth mentioning that for $P_\phi=0$ there is additional work 
\cite{Gossiaux98} showing that Eq.\ (\ref{eq:naive}) is a good 
approximation for any $N$. For simplicity we take $P=1$ for the moment.

The approximation for $c_\phi(\omega, x)$ given by Eq.\ (\ref{eq:naive}) 
allows for an analytical evaluation of $C(x)$ in Eq.\ 
(\ref{eq:temperatureconvolution}) \cite{Efetov95} 
\be 
\label{eq:CdeltaX}
C(\widetilde{x}) = \left(\frac{eT}{\hbar\Gamma}\right)^2 \frac{2\mbox{var}(g_\phi)}
{\widetilde{x}^2 + 1}
\sum_{n=1}^{\infty} \frac{n}{\left(\widetilde{x}^2 + 1 + n 2\pi T/\Gamma
\right)^3},
\ee  

\begin{figure} 
\setlength{\unitlength}{1mm} 
\begin{picture}(90,63)(0,0) 
\put(-5,-60){\epsfxsize=9.0cm\epsfbox{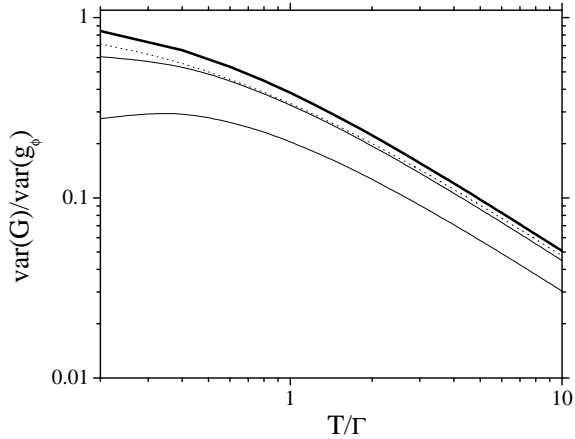}} 
\end{picture} 
\caption{Ratio between var$(G)$ and var$(g_\phi)$ (in units of $e^4/h^2$) 
as a function of $T/\Gamma$.  The sum in Eq.\ (\ref{eq:CdeltaX}) is of
slow monotonic convergence. The thick solid line gives the converged result,
while the two thin ones stand for evaluation of the first two and five 
terms of the sum. 
The dotted line is the approximation given by Eq.\ (\ref{eq:C0fit}).}
\label{fig:scaling+temperature} 
\end{figure}
\begin{figure} 
\setlength{\unitlength}{1mm} 
\begin{picture}(90,72)(0,0) 
\put(-4,-50){\epsfxsize=8.5cm\epsfbox{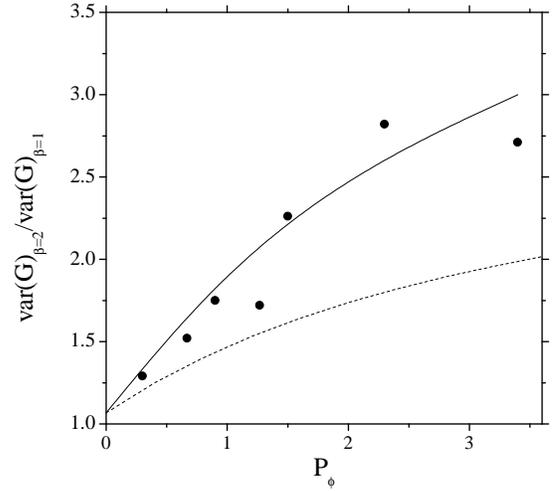}} 
\end{picture} 
\caption{Ratio var$(G)_{\beta=2}$/var$(G)_{\beta=1}$ as a 
function of $P_\phi$. The dashed line stands for the information 
theoretical approach interpolation formula, while the circles 
($\bullet$) are the experimental data, and the solid line is our result. }
\label{fig:variance} 
\end{figure}


\noindent
where $\widetilde{x} = x/X_c$. For $C(0)$ = var($G$), the above expression 
is nicely approximated (within 15\%) by 
\be 
\label{eq:C0fit}
\mbox{var}(G) = \left(\frac{e^2}{h}\right)^2\frac{\mbox{var}\,(g_\phi)}{1 + 2 T/\Gamma} 
\ee 
as shown in Fig. \ref{fig:scaling+temperature}. In Eq. (\ref{eq:C0fit}) the
dependence of var$(G)$ on $\beta$ is implicit in both var($g_\phi$) and
$\Gamma_\phi$. 
Figure \ref{fig:variance} shows that these considerations reconcile 
the theory with the experimental data. 
The information theoretical approach underestimates the ratio
var$(G)_{\beta=2}$/var$(G)_{\beta=1}$ because it lacks the
temperature correction given by $(1 + 2T/\Gamma_{\beta=2})/
(1 + 2T/\Gamma_{\beta=1})$. 
Notice that we do not introduce any additional fitting parameter
in our theory.

\begin{figure} 
\setlength{\unitlength}{1mm} 
\begin{picture}(90,70)(10,0) 
\put(8,-43){\epsfxsize=8.cm\epsfbox{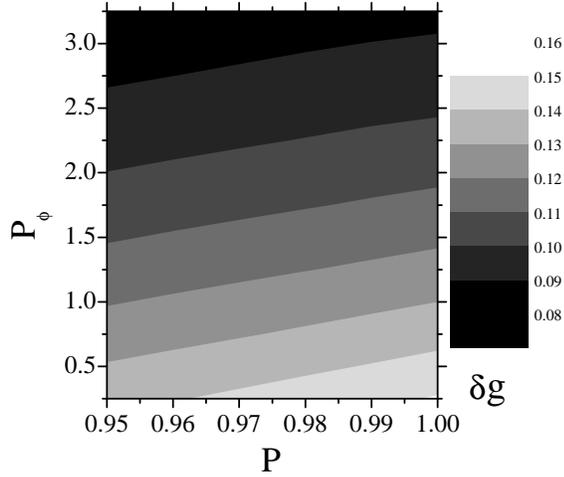}} 
\end{picture} 
\caption{Contour plot of the weak localization peak $\delta g$ as a 
function of the coefficients $P_\phi$ and $P$ for $N=1$.} 
\label{fig:weaklocalization} 
\end{figure}

We now switch our attention to the question of how to determine the 
dephasing rate in open chaotic quantum dots.
There are three main proposed strategies 
\cite{Clarke95,Huibers98b}.
Let us start addressing the one based on the weak localization peak.
As shown  $\langle G \rangle = (e^2/h) \langle g_\phi \rangle$, allows
one to read the average dimensionless conductance directly from the empirical  
data. 
In turn, provided that the leads are ideal ($P=1$), the weak localization 
peak, defined as 
\be 
\delta g = \langle g_\phi \rangle_{\beta=2} - \langle g_\phi \rangle_{\beta=1} 
\ee 
is in  direct relation to $\Gamma_\phi$. \cite{Huibers98b} 
The problem is that in actual experiments $P<1$.
Thus, $\delta g$ is a function not only of $\Gamma_\phi$ but of 
$P$ as well. 
An inspection of Fig. \ref{fig:weaklocalization}, obtained from our
simulations for $N=1$, shows that by 
fixing $\delta g$ (as obtained from the experiment) and decreasing 
$P$ by a small factor always increases $P_\phi$. The effect becomes
small for $P_\phi \gg 1$, but is rather large for $P_\phi \leq 1$.
In this situation, reducing $P$ by 5\% decreases $P_\phi$ by as 
much as 100\%. The dephasing time $\tau_\phi$ increases 
accordingly. Hence, $\delta g$ does not uniquely fixes $P_\phi$.
This ambiguity can by eliminated by using the experimental 
$\langle g_\phi \rangle$ for $\beta =1$ and $2$ to fix both 
coefficients $P$ and $P_\phi$. The data from Ref. \onlinecite{Huibers98a} 
indicate that the correction to $P_\phi$ is significant.

There are two other methods to extract $\tau_\phi$ from experiments 
dealing with chaotic quantum dots. 
Both are related and rely on the study of the parametric dependence
of the conductance. Based on a semiclassical argument \cite{Clarke95}, it was 
proposed that the study of the conductance autocorrelation as a function
of an external magnetic field, $C(\widetilde{x}=B/B_c)$, has a simple 
dependence on $\Gamma_\phi$, namely $(B_c/\Phi_0)^2 = \kappa (2NP + P_\phi)$. 
(Our numerical results support this relation, as depicted in Fig. 
\ref{fig:scaling+dephasing}.) 
By measuring $B_c(T)$ one can thus obtain $\tau_\phi(T)$.
The problem here is that $C(\widetilde{x})$ changes its functional dependence 
when going from the $T\gg \Delta$ to the $T \ll \Delta$ limit \cite{Efetov95}, 
which can jeopardize the determination of $B_c$ for $T \approx \Delta$.
Alternatively, the width of the Lorentzian dip of the average conductance around
$B=0$, can also be used
$\langle g_\phi (B) \rangle= \langle g_\phi \rangle_{\beta=2} -
\delta g/[1 +(B/B_c)^2].
$ 
Both methods were recently shown to give consistent results with the weak 
localization one, at least for $T \ge \Delta$ \cite{Huibers98b}.
This is in apparent contradiction with our claim that there is an ambiguity 
in the weak localization peak method. 
However, differences are only expected for small values of $P_\phi$, where the 
parametric methods were not employed. 
Moreover, since $N=1$ and $P\approx 1$ give $B_c \propto \sqrt{2 + P_\phi}$,
the latter become evidently inaccurate for $P_\phi \ll 1$.

In conclusion, we presented a detailed statistical study of conductance
fluctuations in chaotic quantum dots. We solved the only serious
discrepancy between theory and experiment, giving a stronger support
to the statistical approach incorporating dephasing. In addition, we
pointed out some problems in the quantitative assertion of $\tau_\phi$
from the data, and propose an alternative solution.

E. R. P. Alves acknowledges financial support by CAPES. This work 
was partially funded by CNPq and PRONEX-Brazil.

%%%%%%%%%%%%%%%%%%%%%%%%%%%%%%%%%%%%%%%%%%%%%%%%%%%%%%%%%%%%%%%%%%%%%%%%%%%% 

%%%%%%%%%%%%%%%%%%%%%%%%%%%%%%%%%%%%%%%%%%%%%%%%%%%%%%%%%%%%%%%%%%%%%%%%%% 

\end{multicols}

\end{document}